\documentclass[traditabstract]{aa}
\usepackage{graphicx}
\usepackage{txfonts}
 \usepackage{natbib}
\usepackage{amsmath,amssymb}
\usepackage[modulo, switch]{lineno}
\usepackage[pdftex,pdfnewwindow=true,colorlinks=true,hyperfootnotes=true]{hyperref}
    \hypersetup{colorlinks,%
                 citecolor=blue,%
                 filecolor=blue,%
                 linkcolor=blue,%
                 urlcolor=blue%
      		}

 \bibpunct{(}{)}{;}{a}{}{,} 
  {\left\lbrace\begin{array}{@{}l@{}}}%
  {\end{array}\right.}

\def\be{\begin{equation}}
\def\ee{\end{equation}}    
\def\ba{\begin{eqnarray}}
\def\ea{\end{eqnarray}}

\usepackage{color}

\usepackage{epstopdf}

\begin{document}

\title{A new catalog of magnetically active solar-like oscillators\thanks{Tables~\ref{tab:results} and \ref{tab:results_other} are also available in electronic form at the CDS via anonymous ftp to cdsarc.u-strasbg.fr (130.79.128.5) or via http://cdsweb.u-strasbg.fr/cgi-bin/qcat?J/A+A/}}
\author{E. Corsaro\inst{1}\and
A. Bonanno \inst{1}\and
C. Kayhan \inst{2} \and
M.~P. Di Mauro \inst{3} \and
R. Reda \inst{4,3} \and
L. Giovannelli \inst{4,3} 
          }
\offprints{Enrico Corsaro\\ \email{enrico.corsaro@inaf.it}}

\institute{
    INAF -- Osservatorio Astrofisico di Catania, via S. Sofia, 78, 95123 Catania, Italy
\and
    Department of Astronomy and Space Sciences, Science Faculty, Erciyes University, 38030 Melikgazi, Kayseri, Türkiye
\and
    INAF-IAPS, Istituto di Astrofisica e Planetologia Spaziali, Via del Fosso del Cavaliere 100, 00133, Roma, Italy
\and   
Dipartimento di Fisica, Università degli studi di Roma "Tor Vergata", Via della Ricerca Scientifica 1, 00133, Roma, Italy
}
\date{}

\abstract{We present a new catalog of stars for which detected solar-like oscillations and magnetic activity measurements are both available from chromospheric spectroscopic observations. Our results were obtained by exploiting NASA TESS mission light curves for active stars observed within the Mount Wilson Observatory HK project and the HK survey of the Hamburg Robotic Telescope TIGRE.
We analyzed the light curves for a total of 191 stars by adopting recent techniques based on Bayesian analysis and model comparison to assess the detection of a power excess originating from solar-like oscillations. We characterized the oscillations in a total of 34 targets, for which we provide estimates for the global asteroseismic parameters of $\nu_\mathrm{max}$ (the frequency of maximum oscillation power), $\Delta\nu$ (the large frequency separation), and for the amplitude of the solar-like oscillation envelope $A_\mathrm{max}$. We provide strong statistical evidence for the detection of solar-like oscillations in 15 stars of our sample, identify six further stars where a detection is likely, and 13 stars for which oscillations cannot be ruled out. The key parameters extracted in this work will be exploited for a detailed stellar modeling of the targets and to calibrate relations that connect the level of the measured magnetic activity to the suppression induced on the global oscillation amplitudes. This opens the possibility of shedding light on the interplay between magnetic fields and oscillations. Because of their relatively high brightness, the targets may also be of interest for future dedicated follow-up observations using both photometry and spectropolarimetry.
}
	  
\titlerunning{}
      \authorrunning{}

\maketitle

\section{Introduction}
\label{sec:intro}
Photometric observations of brightness variations induced by stellar oscillations have proven to be a  powerful way for studying stars in detail \citep{Aerts10}. Space missions such as CoRoT\footnote{COnvection ROtation and planetary Transits.} \citep{Baglin06}, NASA \textit{Kepler} and K2 \citep{Boroucki10,Howell14K2}, and TESS\footnote{Transiting Exoplanet Survey Satellite} \citep{Ricker14TESS} have provided and are still providing a wealth of data that can be used to probe fundamental stellar properties such as mass and radius. This enables subsequent modeling aimed at resolving the internal structure and dynamics of stars and at deriving their age. Solar-like oscillations in particular have been observed in tens of thousands of stars with low to intermediate mass because, similarly to the Sun, these oscillations are excited by near-surface convection. This has allowed asteroseismology to consolidate its role over the past years as a key science in fostering our understanding of stars and their evolution. In this context, the study of magnetic fields is of crucial importance because they operate throughout the stellar evolution, starting at the onset of star formation \citep[e.g.][]{McKee07}. As originally found from solar observations \cite{Chaplin00Sun}, magnetic fields impact on stellar oscillations in different ways \cite[e.g.][]{Garcia10,Chaplin11}, but their interplay is not yet fully understood \cite[e.g.][]{Bonanno14,Bonanno19EKEri}. The reason mainly is that it is difficult to find usable targets other than the Sun that are suitable for this type of study, because magnetic activity in stars has the main effect of suppressing the oscillation amplitude \citep[e.g.][]{DiMauro22}. The vast majority of targets with detected solar-like oscillations currently are evolved stars with low levels of magnetic activity, which are clearly not suited for studies of their magnetism. In this respect, increasing the sample of stars with detected oscillations and available measurements of magnetic activity is a mandatory task that needs to be accomplished.

The TESS mission, although not comparable to \textit{Kepler} in terms of photometric precision and duration of the observations, is observing a large number of relatively nearby and bright stars with its large sky coverage. These stars include some for which the magnetic activity index obtained from chromospheric spectroscopy observations of the Ca II H\&K emission lines, for instance, is available. Many targets with measurements of chromspheric activity are part of the historical Mt. Wilson Observatory HK Project \citep[e.g.][]{Wilson78}, to which more were added by means of the HK survey of the Hamburg Robotic Telescope (TIGRE; \citealt{Mittag11TIGRE}). A high fraction of these targets does not fall within the {\it Kepler} field of view of the nominal mission, and therefore, it was not possible to detect potential oscillations through photometric observations before the advent of TESS. 

We exploit the known catalogs of stars with measured magnetic activity from chromospheric spectroscopic observations \citep{Mittag11TIGRE,Marsden14BCool,Boro18cycle,Olspert18cycle} to analyze the corresponding TESS light curves acquired during mission Cycles 1, 2, and 3 in order to search for potential solar-like oscillations. We exploit a Bayesian framework \citep{Corsaro14,Muellner21} to characterize the stellar power spectra and to assess the presence of an oscillation power excess. For the stars with a potential detection, we extract the global asteroseismic properties of $\nu_\mathrm{max}$, $\Delta\nu$, and $A_\mathrm{max}$ using well-consolidated modeling techniques of the stellar power spectra \citep{Kallinger14,Corsaro17metallicity}. We compile a new catalog of stars in this way that will be used to perform subsequent stellar modeling and to investigate the relation between the magnetic activity level and suppression of the oscillation amplitude for a wider range of fundamental stellar properties than was possible before.

\section{Observations and data}
\label{sec:data}
The stars used in this work have been observed by NASA TESS over the past years during Cycles 1, 2, and 3 for a typical duration of about 27 days (one sector) with a cadence of 2~min. By cross-matching the catalogs of active stars with the MAST\footnote{Mikulski Archive for Space Telescopes, \href{https://archive.stsci.edu/}{https://archive.stsci.edu/}.} data archive, we collected usable light curves for a total of 191 stars, 118 of which were taken from \cite{Mittag11TIGRE}, 20 from BCool \citep{Marsden14BCool}, 24 from \cite{Boro18cycle}, and 31 from \cite{Olspert18cycle}.

\begin{figure*}[h!]
    \centering
        \includegraphics[width=18.5cm]{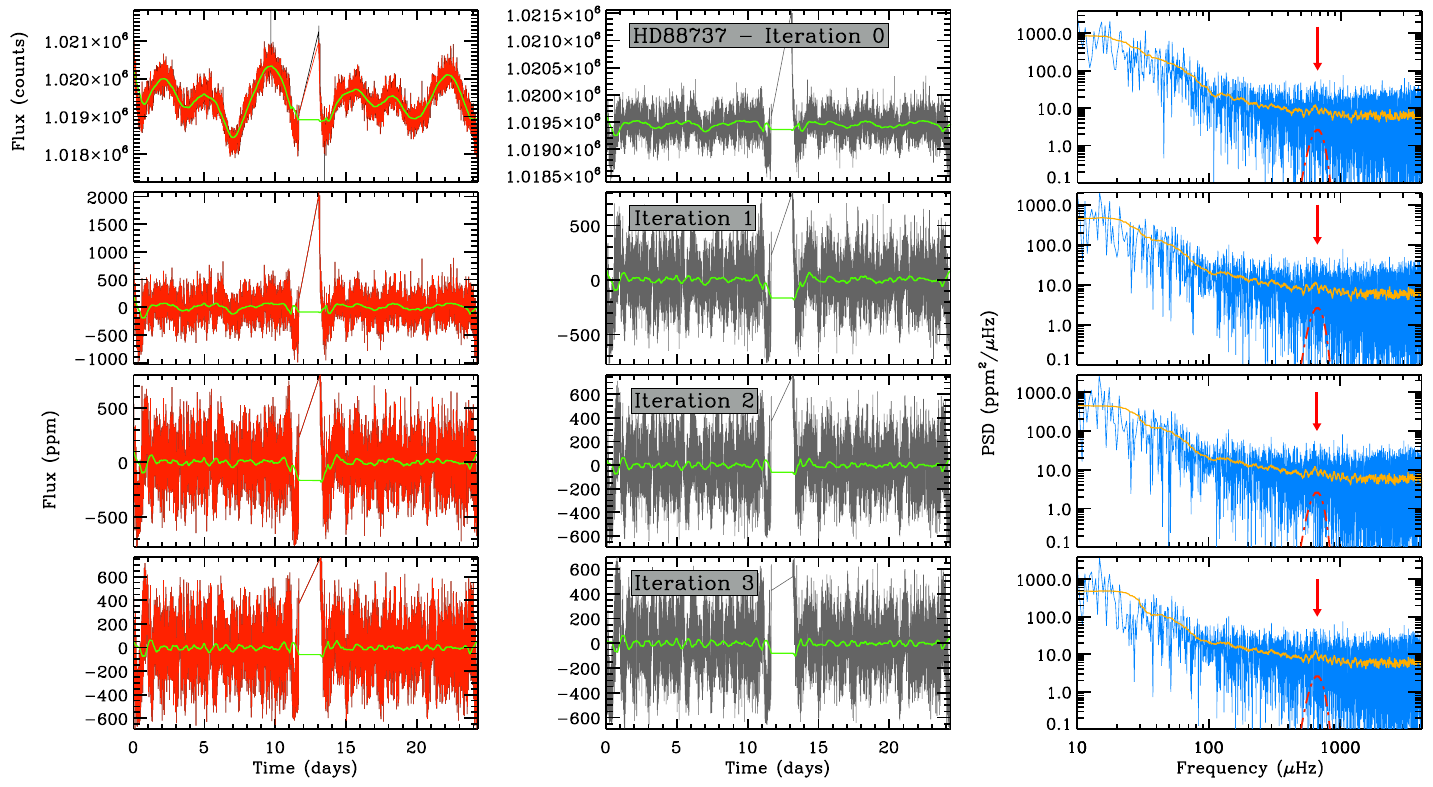}
    \caption{Data preparation process for the light curve of TIC~95340781 (HD~88737). Each row represents a different iteration (four in total), as outlined in Sect.~\ref{sec:data},. The first column shows the light curve before the filtering is applied (in orange), the second column shows the (residual) light curve after the filtering is applied (in gray), and the third column shows the PSD of the residual light curve (in blue). The solid green line shows the smoothing of the light curve at each step, and the yellow line in the third column represents the smoothing of the PSD (i.e., an approximation of the background level; see Sect. \ref{sec:numax}) by a default window of 30 bins. The red arrow marks the position of a power excess typical of solar-like oscillations, here represented by the $\nu_\mathrm{max}$ parameter and the modeled Gaussian function (dot-dashed red line) obtained from the background fit (see Sect.~\ref{sec:astero}). The light-curve units change from counts during the first iteration to ppm for the remainder of the iterations.}
    \label{fig:lc}
\end{figure*}

Using the standard PDCSAP\footnote{Pre-search Data Conditioning Simple Aperture Photometry.} flux as a starting dataset \citep[e.g.][]{Garcia11}, we processed each light curve in order to optimize it for the detection of solar-like oscillations. This was done by first applying a 3$\sigma$ clipping over a shifted light-curve flux with zero mean. Then, a boxcar smoothing was applied, with a varying width adapted to the individual star, and typically ranging from 400 to 800 bins, depending on the level of signal modulation at low frequency. The smoothed light curve was subtracted from the $\sigma$ clipped mean-shifted light curve, and the mean flux of the smoothed light curve was added to the final flux. In addition, the obtained flux was normalized to its median value, shifted to zero, and multiplied by a factor $10^6$ to match the parts-per-million (ppm) units. The procedure involving the 3$\sigma$ clipping and the subtraction of a smoothed light curve was repeated iteratively three more times to improve the removal of long-trend variations that may hamper the detection of oscillations at intermediate to high frequencies due to the leakage of power caused by high-amplitude signal at low frequencies. An example of this processing is shown in Figure~\ref{fig:lc} for the star TIC~95340781 (HD~88737), where an oscillation envelope is detected at $\nu_\mathrm{max} \approx 670 \, \mu$Hz. The first is the most siginficant iteration (first row), allowing the light curve to be flattened out by removing the strongest modulation signal and some prominent outliers (shown in the left panel before the processing was applied, and in central panel after the processing was applied). The second to fourth iterations further increased the flattening of the light curve at shorter timescales, which could be useful to improve the signal-to-noise ratio (S/N) of the oscillation envelope. After the fourth iteration, no further evident improvements were observed, and the procedure was therefore no longer applied beyond this point. The corresponding power spectral density (PSD) is shown in the right panels of the plot, allowing us to keep under control the impact of the filtering on the data product that was subsequently analyzed. We note that the variability shown by the residual light curve on timescales of about a day or shorter typically originates in a real astrophysical signal, such as stellar granulation and oscillations, and therefore, it was intentionally not removed. For reference, the processing of the raw light curve that we described here follows a standard method that is adopted to prepare asteroseismically optimized light curves acquired by NASA \textit{Kepler} \citep{Garcia11}.

To facilitate and guide the subsequent analysis of the PSDs, we collected the available spectroscopic information of each target by exploiting the public PASTEL \citep{Soubiran10PASTEL} and Gaia DR3 catalogs \citep{Gaia,GaiaDR3Apsis}. We considered $\log g$ and $T_\mathrm{eff}$ of the star because they can be used to predict the frequency position of a potential oscillation envelope through the well-known relation that connects $\nu_\mathrm{max}$ to the acoustic cutoff frequency ($\nu_\mathrm{ac}$) \citep{Brown91numax},
\begin{equation}
\label{eq:nu_ac}
    \nu_\mathrm{max} \propto \nu_\mathrm{ac} \propto g / \sqrt{T_\mathrm{eff}} \, ,
\end{equation}
and by scaling this relation to the solar values $\nu_\mathrm{max} = 3150\,\mu$Hz, $T_\mathrm{eff} = 5777\,$K, and $\log g = 4.43775\,$dex \citep{Chaplin11scaling}.
For stars with multiple measurements of $\log g$ and $T_\mathrm{eff}$, we simply considered an average. The predicted values for $\nu_\mathrm{max}$ have the main purpose to represent a guess, or starting point, for the analysis presented in Sect.~\ref{sec:astero}, meaning that we cannot expect these values to necessarily coincide with the final measurements obtained from this work. Because most of the datasets we adopted show high levels of background noise (in particular, photon noise), the reference values for $\nu_\mathrm{max}$ are also useful to guarantee that our potential detections of solar-like oscillations are compliant with the spectroscopic parameters of the star. This helps us to reduce spurious detections (e.g., caused by the contaminating signal from a background star) and to increase the robustness of our results.

\section{Asteroseismic data analysis}
\label{sec:astero}
The analysis of the stellar PSD was carried out in order to estimate $\nu_\mathrm{max}$, detect potential solar-like oscillations, and when this was verified, extract the additional global asteroseismic parameters $\Delta\nu$ and $A_\mathrm{max}$. In the following sections, we describe the details of this analysis.

\subsection{Background fitting and $\nu_\mathrm{max}$}
\label{sec:numax}
After the PSD was obtained as described in Sect.~\ref{sec:data}, it was modeled so that the level of the background signal could be characterized, namely the signal that is not related to solar-like oscillations. This is essential to be able to quantify the S/N of a potential solar-like oscillation envelope, and therefore, to test for its detection. For this purpose, we followed the representation introduced by \cite{Kallinger14}, which comprises the signal from long-trend variations that occur at low frequency (as the result of the combined effects of stellar activity, supergranulation, and other possible instrumental variations), from granulation activity (modeled as a series of Harvey-like profiles; \citealt{Harvey85}), and from photometric white noise. The region containing the solar-like oscillations was instead modeled by a typical Gaussian function. The model we describe here is represented by the following equation:
\begin{equation}
\label{eq:bkg}
    P_\mathrm{PSD} (\nu) = W + \sum_{i=1}^3 \frac{2\sqrt{2}a_i^2/(\pi b_i)}{1 + \left(\nu/b_i\right)^4} + H_\mathrm{osc} \exp{\left[-\frac{\left( \nu - \nu_\mathrm{max} \right)^2}{2 \sigma_\mathrm{env}^2} \right]},
\end{equation}
where $W$ is the instrumental noise, and the number of Harvey-like components can range from one to three, depending on the complexity of the PSD ($a_i$ and $b_i$ represent the amplitude and characteristic frequency, respectively, of the $i$th component), and in turn on the capability of the data to constrain the model. This model clearly permits us to test the hypothesis that the oscillation envelope is present (and therefore, detected) in the given dataset. We followed the method presented by \cite{Corsaro17metallicity} and fit the different background models for every star by means of the  public Bayesian inference software \textsc{Diamonds}\footnote{high-DImensional And multi-MOdal NesteD Sampling, \href{https://github.com/EnricoCorsaro/DIAMONDS}{https://github.com/EnricoCorsaro/DIAMONDS}.} \citep{Corsaro14} and its extension, called \textsc{Background}, which specifically implements the above models (and others as well) for the background fitting in stellar PSDs. This code is particularly suited for this type of application because the fitting parameters are often highly correlated to one another \citep[e.g.][]{Corsaro14}, which makes it more challenging to find a good solution in the parameter space. The high computational and sampling efficiency is made possible by the implementation of a nested-sampling Monte Carlo algorithm \citep{Skilling04}, which has the additional key advantage of providing an estimate of the Bayesian evidence (or marginal likelihood) as a direct output for every model that is fit to the data. The Bayesian evidence is crucial to assess the detection, as we discuss below.

\begin{figure*}[h]
    \centering
        \includegraphics[width=18.6cm]{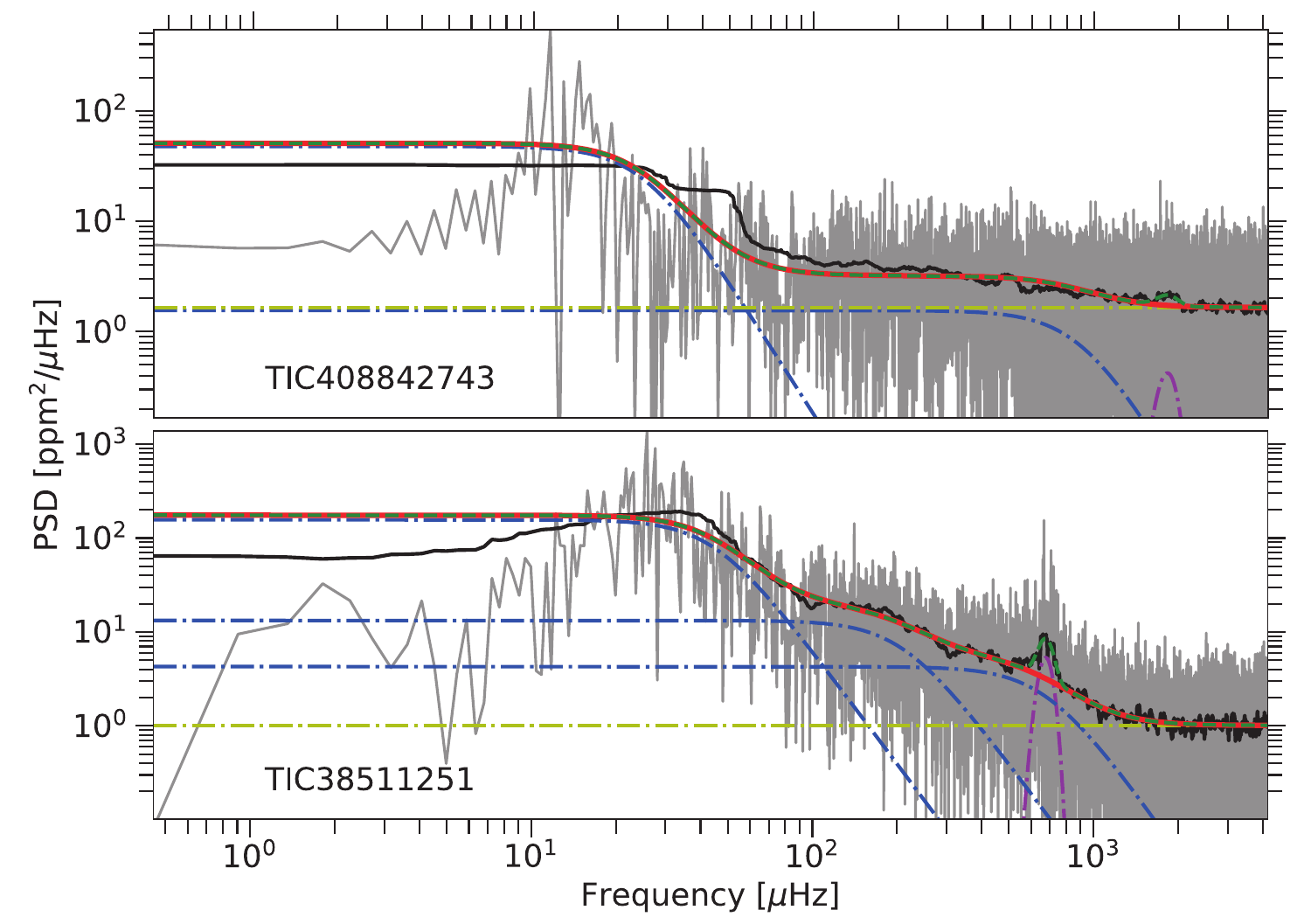}
    \caption{Example of two different background models fit using \textsc{Diamonds} + \textsc{Background}. The models show two Harvey-like profiles for TIC~408842743 and three for TIC~38511251. This results in a larger number of free parameters that need to be estimated in the latter case with respect to the former. The PSD of the star is shown in gray, and its smoothing by $\Delta\nu$ is shown in black. The dot-dashed yellow line represents the level of the white noise, and the Harvey-like components are indicated with dot-dashed blue lines. The dot-dashed magenta line shows the Gaussian envelope of the solar-like oscillations. The overall fit of the background without (with) the Gaussian envelope is marked with a solid red (dashed green) line.}
    \label{fig:bkg}
\end{figure*}

An example of a background fit to the PSD is shown in Figure~\ref{fig:bkg} for the stars TIC~408842743 (HD~187691) and TIC~38511251 (HD~23249), where varying background models were adopted. We chose the adequate model in a Bayesian model comparison framework, as presented by \cite{Corsaro17metallicity}. In practice, this analysis was conducted by first testing the most complex model (i.e., the model with the largest number of Harvey-like profiles) that can be fit by the code to the PSD of a given star. Then, the Bayesian evidence of this model is compared to that of a subsequent model that is slightly simpler (i.e., with one Harvey-like component less than the first model). We therefore selected the most complex model that was tested to determine whether it has the highest Bayesian evidence, and otherwise, we proceeded by including an additional even more simplified model and repeated the model comparison process. We note that when too many Harvey-like profiles were included with respect to what the data were capable of constraining, the code failed to converge to a result. The given model was then discarded because it cannot be reliably included in the model comparison process. In this case, a less complex model was replaced to be tested in addition to the previous model. For most stars, we find that the model accounting for two Harvey-like profiles is favored, followed by the one Harvey-like profile model, and finally, by the three Harvey-like profile model. This is expected because, on the one hand, the instrumental noise level is often so high that the signal in the PSD appears to be nearly flat for most of the observed frequency range (up to the Nyquist frequency), with a power level increasing only toward the low-frequency side. On the other hand, the presence of oscillation envelopes that are suppressed with respect to what we would expect when no magnetic activity is at play further hampers the detection of a second granulation component (the high-frequency component; \citealt{Corsaro17metallicity}), which makes the three Harvey-like profile model less likely and adequate at reproducing the given datasets.

The preliminary step before fitting the model to the data was creating priors for each of the free parameters of the model. This was done by exploiting a public Python routine that is available within the \textsc{Background} code package\footnote{The \textsc{Background} code and related Python routines are available at  \href{https://github.com/EnricoCorsaro/Background}{https://github.com/EnricoCorsaro/Background}.}. In particular, the routine in charge of building up the prior distributions operates as follows: \textit{i)} It evaluates a median signal in a frequency range close to the Nyquist frequency to obtain a guess on the level of the photometric white noise $W$. \textit{ii)} It uses the input guess of $\nu_\mathrm{max}$ found from the PASTEL catalog (by combining $\log g$ and $T_\mathrm{eff}$ as in Eq.~\ref{eq:nu_ac}) to evaluate the priors on amplitude and characteristic frequency of the Harvey-like components of the granulation activity, as shown by \cite{Kallinger14} and by \cite{Corsaro17metallicity}. \textit{iii)} It obtains the maximum level of the signal at very low frequency (below the characteristic frequency of any subsequent Harvey-like profile corresponding to granulation) to provide an estimate of the amplitude for the low-frequency component, which is not directly related to the granulation activity of the star. Additionally, the priors on the Gaussian envelope of the solar-like oscillations were generated by using the input $\nu_\mathrm{max}$ to compute an estimate of the standard deviation of the envelope through the empirical relation \citep[e.g.][]{Mosser12}, and an estimate of $\Delta\nu$ obtained from the scaling relation \citep[e..g.][]{Stello09,Huber11}, which was used to evaluate a smoothing of the PSD by a window proportional to $\Delta\nu$. This smoothed PSD was then adopted to determine the highest power in the range $\nu_\mathrm{max} \pm 3\Delta\nu$, so that a prior on $H_\mathrm{osc}$ can be set. All prior distributions were made uniform, which is found to provide better computational speed during the nested-sampling Monte Carlo process than using more complex distributions.

\subsection{Bayesian detection of solar-like oscillations}
\label{sec:bayes}
After this first set of fits was carried out for every star in the sample, we proceeded with the evaluation of the detection of the oscillation power excess. We followed a procedure along the lines presented by \cite{Corsaro14} and \cite{Muellner21}, who adopted the same fitting code. In addition to the first set of background models described above, we considered a new set of background models that contained the same components as those that were fit at the beginning, except that no Gaussian envelope was included. This new set of models therefore allowed us to test the hypothesis that oscillations are absent, or are at least not detected, in the observed stellar PSD. We used the same input prior distributions as we adopted for the first set of fits, but now removed those related to the parameters of the Gaussian envelope. When the fits for the second set of models were also finalized, we performed the Bayesian model comparison by evaluating the Bayes factor as follows:
\begin{equation}
    \ln \mathcal{B}_{1,0} = \ln \mathcal{E}_1 - \ln \mathcal{E}_0 \, ,
\end{equation}
where $\mathcal{E}_1$ is the Bayesian evidence computed by fitting the model incorporating the Gaussian envelope of the solar-like oscillations (i.e., the detection hypothesis), and $\mathcal{E}_0$ is the Bayesian evidence when the Gaussian envelope is not included (i.e., the nondetection hypothesis). We followed the standard Jeffreys empirical scale of strength for the evidence \citep{Jeffreys61}, which suggests that only when $\ln \mathcal{B}_{1,0} \geq 5.0$  can we claim a clear detection of the oscillation signal, while falling into regimes of moderate to weak detection for $1 \leq \ln \mathcal{B}_{1,0} < 5$ 
\citep[see also][]{Trotta08}, meaning that there is still a chance that power excess is present, but more data are needed to confirm this. We present our results in Sect.~\ref{sec:results}

\subsection{Measuring $\Delta\nu$}
\label{sec:dnu}
This part of the analysis aims to measure the global asteroseismic parameter $\Delta\nu$. This analysis can be conducted only after a solar-like oscillation envelope is found. In our case, we applied this additional analysis not only to all stars for which a detection provided a strong evidence condition, but also more in general for  stars for which an oscillation envelope was either still likely or could not be completely ruled out.

\begin{figure*}[h]
    \centering
        \includegraphics[width=18.5cm]{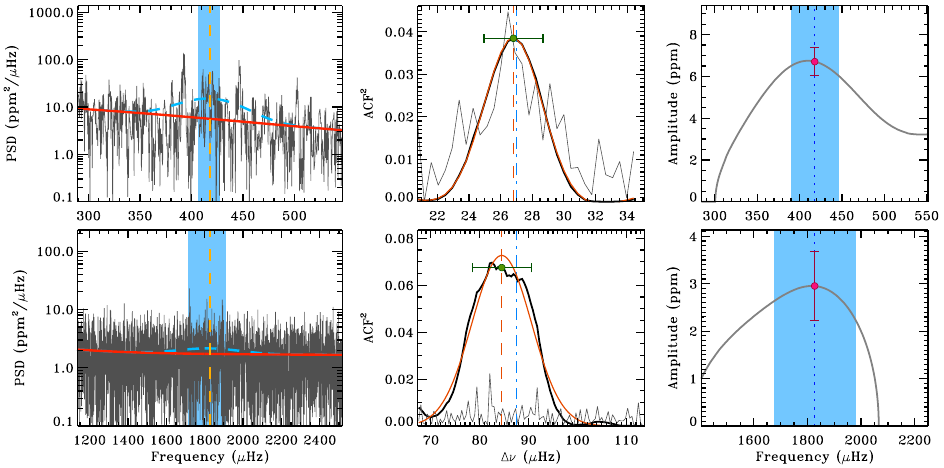}
    \caption{Example of measuring $\Delta\nu$ and $A_\mathrm{max}$ for the stars TIC~375621179 (HD~188512, top) and TIC~408842743 (HD~187691, bottom). \textit{Left panels}: Stellar PSD (in gray). The background fit without the Gaussian envelope is overlaid (in red). The fitted Gaussian envelope above the background is marked with a dashed blue line. The vertical blue band corresponds to a $3$ $\sigma$ Bayesian credible region around $\nu_\mathrm{max}$ (vertical dashed orange line) as obtained from \textsc{Diamonds}. \textit{Middle panels}: ACF$^2$ function to estimate $\Delta\nu$ as computed on the original PSD (light gray curve) and on the PSD smoothed by $\Delta\nu/10$ (thick black curve). The dot-dashed blue line marks the position of $\Delta\nu$ from the scaling relation. The selected Gaussian fit is shown as the orange curve, and its centroid is indicated by the vertical dashed orange line. The final estimate of $\Delta\nu$ is represented by the green bullet, and its 1$\sigma$ uncertainty is overlaid. \textit{Right panels}: Smoothed amplitude spectrum (in gray). The estimate of $A_\mathrm{max}$ and its 1$\sigma$ uncertainty are marked by the magenta bullet at the position of $\nu_\mathrm{max}$ (vertical dotted blue line). The shaded blue region depicts the range over which the uncertainty on $A_\mathrm{max}$ is computed, $\nu_\mathrm{max} \pm \sigma_\mathrm{env}$.}
    \label{fig:dnu}
\end{figure*}

To estimate $\Delta\nu$, we relied on a classic method that enabled us to identify a regular structure in the data, namely the autocorrelation function (ACF) \citep[e.g. see][]{Mosser09}. The middle panels of Figure~\ref{fig:dnu} show an example of this application for the stars TIC~375621179 and TIC~408842743. The searching range for the estimation of $\Delta\nu$ was considered to be $\pm\,30$\% of an initial $\Delta\nu$ guess based on $\nu_\mathrm{max}$, which in turn was evaluated using the $\Delta\nu$ -- $\nu_\mathrm{max}$ relation provided by \cite{Huber11}. The region of the PSD we used to evaluate the ACF was centered around $\nu_\mathrm{max}$. In principle, this region should be set to be as large as possible so that it encompasses a larger number of oscillations modes, but at the same time, it should not be too large to avoid including extended low-S/N regions that can hamper the ACF signal. We therefore compared the extension of the range when considering either $\nu_\mathrm{max} \pm 2 \sigma_\mathrm{env}$ or $\nu_\mathrm{max} \pm 3 \Delta\nu$, and for each star, we took the larger of the two. This is because when $\sigma_\mathrm{env}$ is at least two times $\Delta\nu$, the oscillation envelope spans over several radial orders, and it is therefore more likely that the chances to identify a stronger ACF signal are increased. The ACF$^2$ was then fit in two steps using a nonlinear least-squares fit each time. In the first step, we adopted a combination of a Gaussian and a quadratic function (from four to six terms), and in the second step, we used the results from the first step to set up the starting points for a second fit using just a simple Gaussian function (three terms only, hence without any quadratic component). We took as the best fit the fit whose Gaussian had a centroid that was closest to the position of the maximum of the ACF$^2$. The final estimate of $\Delta\nu$ was therefore the centroid of the best-fit Gaussian function, with a corresponding uncertainty obtained from the same fit.

To provide an additional level of support to the analysis presented here and in Sect.~\ref{sec:numax}, we adopted another method for extracting the global asteroseismic properties of $\nu_\mathrm{max}$ and $\Delta\nu$. As a check, we applied this different method to the stars belonging to the TIGRE catalog, which is the largest catalog of those adopted for the sample selection in Sect.~\ref{sec:data}. In this case, we used light curves obtained with the \texttt{lightkurve} code package \citep{2018ascl.soft12013L}. In order to determine $\nu_\mathrm{max}$, we applied a 2D autocorrelation function (ACF2D) method \citep{2009CoAst.160...74H, 2019ApJ...879...33V} to the stellar power spectrum according to varying window widths (10-100 $\mu$Hz) and spacing parameters. In this method, we gradually applied a shift to the data to find an optimal correlation. To diagnose the $\nu_\mathrm{max}$ values, we checked the highest S/N oscillation frequency lag and also the correlation metric in the ACF2D. We determined the central oscillation frequency of each segment that was
related to the strength of the ACF2D in the oscillation frequency lag. As the ACF2D signal increases, the maximum oscillation frequency finds the best value. Then, we determined the peak of the ACF2D and considered the center of the Gaussian function fit to the peak as the final $\nu_\mathrm{max}$ value. In the 
$\nu_\mathrm{max}$ region, we further probed the ACF2D to possibly estimate $\Delta\nu$. 
Depending on $\nu_\mathrm{max}$, the full width at half maximum (FWHM) of the mode envelope was shaped. Using well-established empirical relations in \citet{2010A&A...517A..22M} and \citet{2017ApJ...835..172L}, we collapsed the calculated ACF2D region to determine an approximate $\Delta\nu$ in a manner similar to the method used in  $\nu_\mathrm{max}$. It is well known that the ACF2D is sensitive to regularly shaped peaks. Therefore, we fit a Gaussian function to the peak between the nearest empirical $\Delta\nu$ to determine the observed $\Delta\nu$. For the $\nu_\mathrm{max}$ and $\Delta\nu$ uncertainties, we applied 
the method of \citet{2009CoAst.160...74H}. For high-S/N levels, the uncertainties of $\nu_\mathrm{max}$ 
and $\Delta\nu$ are above 10 $\mu$Hz and 0.1 $\mu$Hz, respectively. A comparison of the measurements obtained from the Bayesian background fitting presented in Sect.~\ref{sec:bayes} (for $\nu_\mathrm{max}$) and the ACF analysis presented at the beginning of this section (for $\Delta\nu$), with those obtained from the ACF2D method (for both $\nu_\mathrm{max}$ and $\Delta\nu$), yields an agreement of $\sim 4\,$\% for $\nu_\mathrm{max}$ and $\sim 8$\,\% for $\Delta\nu$ on average. This agrees with the uncertainties obtained in this work.

\subsection{Measuring $A_\mathrm{max}$}
\label{sec:amax}
The last part of the analysis is related to the estimation of the global oscillation amplitude $A_\mathrm{max}$. The amplitudes of solar-like oscillations are known to be related to the fundamental properties of stars, such as mass, luminosity, and temperature \citep[e.g.][]{Kjeldsen95,Corsaro13}. However, since oscillation amplitudes originate from the interplay between the excitation and damping mechanism, they are subject to stochastic effects. To properly measure the global oscillation amplitudes, we therefore relied on the recipe developed by \cite{Kjeldsen05,Kjeldsen08}. This involves the smoothing of the PSD by a Gaussian FWHM of $4\Delta\nu$, the subtraction of the background level, and the rescaling by $c/\Delta\nu$, with $c = 3.04$ to obtain the power per radial mode. An example of the result of this analysis is shown in the right panel of Figure~\ref{fig:dnu}. After the smoothed amplitude spectrum was obtained, $A_\mathrm{max}$ was evaluated as the amplitude at the position of $\nu_\mathrm{max}$. We note that this value may not necessarily coincide with the absolute maximum of the smoothed amplitude curve. As an uncertainty in $A_\mathrm{max}$, we considered the largest variation of $A_\mathrm{max}$ found within the range $\nu_\mathrm{max} \pm \sigma_\mathrm{env}$, which is representative of the breadth of the oscillation envelope region. Our rationale stems from the hypothesis that a flatter oscillation envelope, indicative of a lower S/N, might result in a less constrained $A_\mathrm{max}$ compared to narrower, more peaked envelopes. The extent and flatness of this envelope assume significance because a pronounced suppression effect is induced by the magnetic fields. Our uncertainty therefore reflects how much $A_\mathrm{max}$ is subject to change within the central region of the oscillation envelope. Furthermore, our delineation of the uncertainty is inherently linked to $\nu_\mathrm{max}$, as expected because $A_\mathrm{max}$ is by definition evaluated at the position of $\nu_\mathrm{max}$. At the same time, this uncertainty represents a more conservative estimate than is derived from only mapping the Bayesian credible region estimated for the parameter $\nu_\mathrm{max}$. We note that this is a different definition of the uncertainty than was given by \cite{Huber11}, who relied on Monte Carlo simulations to inject different realizations of the photometric noise. This latter definition inherently underestimates the uncertainty as it overlooks potential effects such as variability induced by the stellar magnetic cycle, stochastic oscillations, and fluctuations in granulation amplitude.

Nonetheless, it is crucial to note that while the S/N and flatness of the oscillation envelope contribute significantly to the level of statistical significance determined through the model comparison (as detailed in Sect.~\ref{sec:bayes} and Tables~\ref{tab:results} and ~\ref{tab:results_other}), other factors concurrently influence this determination. They include the frequency range of the oscillation region, the shape of the background signal below the oscillation envelope, potential deviations of the oscillation envelope from an ideal Gaussian bell curve, and the number of data bins within the modeled region. Additionally, the as-yet-uncalibrated effect of suppression induced by magnetic activity in this specific TESS target sample \citep[e.g.][]{Chaplin11} further complicates this assessment. This implies that a more precise estimate of $A_\mathrm{max}$ is not necessarily associated with a stronger detection of the oscillation envelope. A detailed study of the reliability of the adopted uncertainties in $A_\mathrm{max}$ is envisioned for a subsequent work, where the amplitudes will be analyzed in the context of the suppression effect induced by the magnetic activity.

\section{Results}
\label{sec:results}
For the sake of completeness, in addition to the stars for which a detection of the oscillation envelope is highly likely (i.e., $\ln \mathcal{B}_{1,0} \ge 5$), we decided to report all the stars resulting in $\lvert \ln \mathcal{B}_{1,0} \rvert < 5$, meaning that the detection is either a likely scenario or that at least it cannot be totally ruled out. The value of the Bayes factor for each star in the sample therefore provides a statistical weight on the detection, so that the user can decide whether to rely on the result, depending on the needs. All targets for which a power excess is not favored (i.e., $\ln \mathcal{B}_{1,0} < 0$) still exhibit a potential indication of an oscillation envelope that future observations may eventually confirm or disprove. We list all the results in Table~\ref{tab:results} for the stars for which we find a strong detection condition, meaning that they satisfy the Bayesian model selection process because they show a clear and prominent oscillation power excess, and in Table~\ref{tab:results_other}, we list all the other stars for which we were unable to draw a firm conclusion. The stars are listed by increasing HD identifier for each catalog, where the asteroseismic parameters of $\nu_\mathrm{max}$, $\Delta\nu$, and $A_\mathrm{max}$ are obtained as outlined in Sect.~\ref{sec:astero}. Moreover, Tables~\ref{tab:results} and~\ref{tab:results_other} appear to show that many of the stars analyzed in this work belong to the catalog source of \cite{Olspert18cycle}, for which a magnetic activity cycle period and rotation period measurements are also available. This makes these targets even more suitable and interesting for future dedicated studies.

Because of the low S/N that is typically found in the datasets analyzed in this work, the fitting of the background models was a challenging task. In order to make \textsc{Diamonds} + \textsc{Background} converge to a stable solution, it was mandatory to perform the fits multiple times for many of the targets. This was done to improve the choice of the initial prior distributions in a way that allows the likelihood maximum in the parameter space of the solutions to be fully resolved. This was conducted by means of a hierarchical approach, using the former posterior solution as input to refine the prior distributions of a subsequent inference. 

Despite the low S/N of the data, the measurement of $\Delta\nu$ through the computation of the ACF$^2$ produced a clear signal of a comb-like pattern, that is, reliable results in most stars, well in line with the prediction obtained from the scaling relation \citep{Huber11} (within just a few percent). The middle panels of Figure~\ref{fig:dnu} clearly show that the ACF$^2$ on the smoothed PSD performs significantly better than that applied on the original PSD. This is because when smoothing the PSD by some fraction of $\Delta\nu$ (in this case, $\Delta\nu/10$), most of the stochasticity that characterizes this type of data can easily be removed, resulting in a net improvement in showing the comb-like structure that is typical of the solar-like oscillations. 

Our recipe for measuring $A_\mathrm{max}$ is not absolute; other definitions may be used. However, the way of measuring the global oscillation amplitudes described in Sect.~\ref{sec:astero} has commonly been adopted in the literature, even for observations based on radial velocity measurements rather than photometric ones. This clearly allows us to rely on a standard reference definition, which can be used to place our new measurements into context with previous ones, such as those published by \cite{Huber11} for a large number of \textit{Kepler} targets, in particular, for the stars observed within the Sounding Stellar Cycles with \textit{Kepler} program by \cite{Karoff09} \citep[see also][]{Karoff13,Bonanno14}, comprising targets with both solar-like oscillations and measured magnetic activity.

We note that for the star TIC~118572803 (HD~22049), which has the highest $\nu_\mathrm{max}$ of the sample, the analysis of $\Delta\nu$ and $A_\mathrm{max}$ was more challenging than for other stars. It was not possible to constrain $A_\mathrm{max}$ to better than a mere 100\,\% precision. Even $\Delta\nu$ appears to deviate with respect to the scaling relation value (it is found to be below it by more than 10\,\%). We ascribe these discrepancies to two main aspects, namely the fact that $\nu_\mathrm{max}$ of this star is close to the Nyquist frequency of the dataset, which was set to about $4167\,\mu$Hz, and therefore does not permit fully resolving the extension of the potential power excess, and the very low S/N of the expected asteroseismic signal, which in turn can be attributed to the combined effect of magnetic activity and small oscillation amplitudes that are typical of a star in this regime of high oscillation frequencies \citep[e.g.][]{Corsaro13}. As expected, the Bayesian model comparison for this star suggests an inconclusive detection process, so that the reported result has to be considered with caution.

\begin{table*}[h]
    \caption{Global asteroseismic parameters $\nu_\mathrm{max}$, $\Delta\nu$, and $A_\mathrm{max}$ and their corresponding 1$\sigma$ uncertainties as obtained in this work. For each star, the TESS Input Catalog and HD ID are provided, along with the reference catalog of the magnetic activity that was used. The last column displays the Bayes factor for the detection of solar-like oscillations, showing only stars with confirmed detections, obtained for $\ln \mathcal{B}_{1,0} \ge 5$.}
    \centering
    \begin{tabular}{llllccr}
    \hline
    HD &  TIC &  Catalog & $\nu_\mathrm{max}$ ($\mu$Hz) & $\Delta\nu$ ($\mu$Hz) & $A_\mathrm{max}$ (ppm) & $\ln \mathcal{B}_{1,0}$\\
\hline
\hline
3795       &      20926643      &    TIGRE                  &        $ 815_{-10}^{+9}$     &     $48.34 \pm 4.67$      &     $7.0 \pm 0.4$ &    $32.90$ \\
23249      &      38511251      &    TIGRE                  &        $ 676_{-5}^{+5}$     &     $ 40.63   \pm      2.52$      &     $ 6.0    \pm     0.6$ &    $72.43$  \\  
117176     &      95473936      &    TIGRE                  &        $ 946_{-17   }^{+13}$     &     $ 52.22   \pm      2.71$      &     $ 5.1    \pm     1.4$ &    $ 22.76   $  \\  
121370     &      367758676     &    TIGRE                  &        $ 686_{-     26   }^{+    27}$     &     $ 39.16   \pm      2.60$      &     $12.7    \pm     4.8$ &    $ 25.57   $  \\  
142373     &      157364190     &    TIGRE                  &        $1036_{-     13   }^{+    17}$     &     $ 58.35   \pm      2.88$      &     $ 4.2    \pm     0.7$ &    $ 39.39   $  \\  
219834A    &      214664574     &    TIGRE/Boro/Olspert     &        $ 816_{-     20   }^{+    14}$     &     $ 50.82   \pm      2.64$      &     $ 5.5    \pm     1.5$ &    $ 52.87   $  \\  
4628       &      257393898     &    Boro/Olspert           &        $2631_{-     70   }^{+    49}$     &     $119.90   \pm      4.20$      &     $ 4.0    \pm     1.1$ &    $  5.96  $  \\   
81809      &      46802551      &    Boro/Olspert           &        $ 716_{-     12   }^{+    11}$     &     $ 49.52   \pm      1.68$      &     $ 3.0    \pm     0.3$ &    $ 15.19   $  \\
27022      &      84982715      &    Olspert                &        $  92_{-      2   }^{+     3}$     &     $  7.48   \pm      0.06$      &     $10.3    \pm     1.4$ &    $  5.00  $  \\   
68290      &      125056469     &    Olspert                &        $ 105_{-      2   }^{+     2}$     &     $ 10.17   \pm      0.18$      &     $16.0    \pm     1.7$ &    $ 17.59   $  \\ 
88737      &      95340781      &    Olspert                &        $ 668_{-     18   }^{+    20}$     &     $ 34.10   \pm      1.23$      &     $ 4.7    \pm     0.8$ &    $  9.58  $  \\  
187691     &      408842743     &    Olspert                &        $1827_{-     38   }^{+    27}$     &     $ 84.51   \pm      6.07$      &     $ 3.0    \pm     0.7$ &    $ 10.86   $  \\  
188512     &      375621179     &    Olspert                &        $ 418_{-      4   }^{+     3}$     &     $ 26.82   \pm      1.87$      &     $ 6.7    \pm     0.7$ &    $ 56.84   $  \\  
218658     &      354379201     &    Olspert                &        $ 129_{-      2   }^{+     2}$     &     $ 10.04   \pm      0.49$      &     $20.3    \pm     3.1$ &    $ 78.34   $  \\  
35296      &      47346402      &    BCool                  &        $1911_{-     27   }^{+    25}$     &     $ 97.07   \pm      3.75$      &     $ 5.5    \pm     0.7$ &    $ 16.77   $  \\   
\hline
    \end{tabular}
    \label{tab:results}
\end{table*}

\begin{table*}[h]
    \caption{Same as for Table~\ref{tab:results_other}, but for stars with moderate, weak, and inconclusive detections ($\lvert \ln \mathcal{B}_{1,0} \rvert < 5$).}
    \centering
    \begin{tabular}{llllccr}
    \hline
    HD &  TIC &  Catalog & $\nu_\mathrm{max}$ ($\mu$Hz) & $\Delta\nu$ ($\mu$Hz) & $A_\mathrm{max}$ (ppm) & $\ln \mathcal{B}_{1,0}$\\
\hline
\hline
9562       &      29845542      &    TIGRE                  &        $1201_{-40}^{+    51}$     &     $ 61.25   \pm      2.68$      &     $ 2.0    \pm     1.0$ &    $ -3.41  $ \\   
114710     &      445070560     &    TIGRE                  &        $2730_{-    108   }^{+   108}$     &     $130.99   \pm      4.80$      &     $ 4.5    \pm     1.3$ &    $  4.55  $  \\   
118972     &      101969923     &    TIGRE                  &        $ 783_{-     43   }^{+    44}$     &     $ 54.88   \pm      2.41$      &     $ 4.9    \pm     0.9$ &    $ -2.87  $  \\   
120136     &      72506701      &    TIGRE                  &        $1139_{-     21   }^{+    32}$     &     $ 67.37   \pm      3.76$      &     $ 2.1    \pm     0.2$ &    $ -1.30  $  \\   
2454       &      466443343     &    Olspert                &        $ 951_{-     26   }^{+    22}$     &     $ 54.28   \pm      2.69$      &     $ 3.9    \pm     0.8$ &    $  1.11  $  \\   
10072      &      327346683     &    Olspert                &        $  69_{-      6   }^{+     5}$     &     $  6.85   \pm      0.38$      &     $13.3    \pm     2.3$ &    $  1.29  $  \\   
22049      &      118572803     &    Olspert                &        $3767_{-    159   }^{+    87}$     &     $141.08   \pm      7.26$      &     $ 3.2    \pm     3.2$ &    $  0.54 $  \\    
57727      &      26693841      &    Olspert                &        $ 119_{-      4   }^{+     4}$     &     $ 10.99   \pm      0.47$      &     $ 7.4    \pm     1.1$ &    $  2.83  $  \\   
76572      &      166529988     &    Olspert                &        $ 826_{-     81   }^{+    66}$     &     $ 51.13   \pm      1.54$      &     $ 2.2    \pm     0.4$ &    $ -2.89  $  \\   
82635      &      188511154     &    Olspert                &        $  100_{-     10   }^{+     9}$     &     $  8.20   \pm      0.83$      &     $ 8.6    \pm     1.5$ &    $  1.41  $  \\ 
85444      &      396811649     &    Olspert                &        $  62_{-      1   }^{+     1}$     &     $  4.72   \pm      0.43$      &     $11.7    \pm     3.5$ &    $ -1.68  $  \\     
88373      &      26208724      &    Olspert                &        $ 535_{-     23   }^{+    23}$     &     $ 32.67   \pm      2.15$      &     $ 6.4    \pm     1.3$ &    $ -3.24  $  \\ 
161239     &      460022840     &    Olspert                &        $ 415_{-     10   }^{+     9}$     &     $ 24.03   \pm      0.66$      &     $21.5    \pm     4.6$ &    $ -1.56  $  \\   
206860     &      301880196     &    Olspert                &        $2703_{-     83   }^{+    75}$     &     $118.85   \pm      2.69$      &     $ 4.8    \pm     1.4$ &    $ -3.70  $  \\  
4614       &      445258206     &    BCool                  &        $2868_{-    100   }^{+    71}$     &     $126.04   \pm      5.18$      &     $ 2.1    \pm     0.5$ &    $ -0.38 $  \\   
13043      &      250419029     &    BCool                  &        $1878_{-     75   }^{+    45}$     &     $ 91.02   \pm      1.67$      &     $ 4.3    \pm     1.1$ &    $ -2.58  $  \\  
30562      &      176379354     &    BCool                  &        $1499_{-     42   }^{+    83}$     &     $ 74.52   \pm      4.22$      &     $ 2.8    \pm     0.9$ &    $ -0.01$  \\  
30652      &      399665349     &    BCool                  &        $2992_{-     76   }^{+   108}$     &     $114.93   \pm      6.81$      &     $ 2.5    \pm     1.7$ &    $ -1.14  $  \\    
95128      &      21535479      &    BCool                  &        $2473_{-     73   }^{+   146}$     &     $103.25   \pm      3.69$      &     $ 2.3    \pm     0.5$ &    $ -2.08  $  \\
\hline
    \end{tabular}
    \label{tab:results_other}
\end{table*}

\section{Discussion and conclusion}
The analysis conducted in this work enabled us to derive global asteroseismic properties of $\nu_\mathrm{max}$, $\Delta\nu$, and $A_\mathrm{max}$ for a sample of 34 stars exhibiting measurable levels of magnetic activity, which are available from the catalogs of \cite{Mittag11TIGRE,Marsden14BCool,Boro18cycle,Olspert18cycle}, comprising also observations from the Mount Wilson HK project (except for HD~118972). Fifteen of these 34 stars show strong evidence for the detection of a solar-like oscillation envelope, while for 6 further targets, our analysis suggests that the presence of oscillations is a likely scenario, but it cannot be confirmed with the current datasets. For an additional 13 stars, we cannot rule out that the oscillation envelope is not detected ($\ln B_{1,0} \leq 0$), and therefore, we decided to include these targets in our compiled list.

Tables~\ref{tab:results} and~\ref{tab:results_other} show that the sample we analyzed covers stars from the main sequence up to the red giant branch, that is, the sample spans a considerable range in stellar evolution. We note that despite the 27\,days of observation, the highest S/N stars (corresponding to stars with the highest values of $\ln \mathcal{B}_{1,0}$) exhibit a PSD that can be further analyzed in terms of detailed oscillation properties, that is, individual mode oscillation frequencies, line widths, and amplitudes, through a dedicated peak bagging analysis \citep[see e.g.][]{Corsaro15cat,Corsaro20FAMED}. This guarantees an additional level of accuracy in the obtained modeling results and opens the possibility of investigating the impact of magnetic activity on the individual oscillation mode parameters. 

The precision attained on our inferred global asteroseismic properties is adequate for a subsequent analysis to investigate how these properties relate to the measured level of magnetic activity. This will be done in a subsequent work currently in preparation through the inclusion of the stellar ages and of the suppression of the oscillation amplitudes. Following the work done by \cite{Bonanno14, Bonanno22}, the results provided here set the basis for a dedicated modeling of the targets with the aim to understand how stellar ages (which will be obtained through an asteroseismic-based inference) are connected to the level of oscillation amplitude suppression, rotation, and metallicity, that is, of magnetic activity in stars. 

Future missions such as ESA PLATO \citep{Rauer14PLATO} will enable a significant extension of this type of studies to stars covering a wider range of fundamental stellar properties, and they will likely increase the final outcome by about two orders of magnitude already through the inclusion of stars belonging to the P1 and P2 (primary) samples of bright targets. Solar-like oscillations constitute a key source of information that enable us to infer the internal structure of stars. This work in conjunction with the analysis of the interplay between magnetic activity cycles and the dynamical properties of rotation and convection \citep[e.g.][]{Bonanno22} therefore represents a main path to follow for the years to come in order to shed light on the role of magnetic activity in the stellar evolution. Finally, the stars provided here are all relatively bright stars, and therefore, additional follow-up studies involving spectropolarimetry and ground-based photometry observations might be possible in principle. This will enable the inference of the magnetic field topology and will thus contribute to improve our understanding of the behavior and role of magnetic fields in stars with solar-like oscillations.

\section*{Acknowledgments}
E.C. and A.B. acknowledge support from PLATO ASI-INAF agreement no. 2022-28-HH.0 "PLATO Fase D". We acknowledge support from the research grant “Unveiling the magnetic side of the Stars” (PI A. Bonanno) funded under the INAF national call for Fundamental Research 2023.
\bibliographystyle{aa}
\bibliography{biblio}
\end{document}